\documentclass[aps,pre,twocolumn,amsmath,amssymb,showpacs,amsfonts]{revtex4}
\usepackage{epsfig}
\usepackage{graphicx}
\usepackage{dcolumn}
\usepackage{bm}

\begin{document}

\title{Conformal Field Theory as Microscopic Dynamics of Incompressible\\
Euler and Navier-Stokes Equations}

\author{Itzhak Fouxon}
\author{Yaron Oz}

\affiliation{Raymond and Beverly Sackler School of Physics and Astronomy,
Tel-Aviv University, Tel-Aviv 69978, Israel}

\date{\today}

\begin{abstract}

We consider the hydrodynamics of relativistic conformal field theories at finite temperature. We show that the limit of slow motions
of the ideal hydrodynamics leads to the non-relativistic
incompressible Euler equation. For viscous hydrodynamics
we show that the limit of slow motions
leads to the non-relativistic incompressible Navier-
Stokes equation.
We explain the physical
reasons for the reduction and discuss the implications.
We propose that
conformal field theories provide a fundamental
microscopic viewpoint of the equations and the dynamics governed by them.

\end{abstract}

\pacs{11.25.Hf,47.10.ad,11.25.Tq}

\maketitle

Turbulence is a central unsolved problem of classical physics. The main obstacle to an
advancement in the area is the insufficient understanding of the equations governing the non-linear
dynamics of the fluids. In a most basic situation  these are the
incompressible Navier-Stokes (NS) equations (see e.g \cite{Frisch})
\begin{eqnarray}&&
\partial_t \bm v+(\bm v\cdot\nabla)\bm v=-\nabla P+\nu\nabla^2 \bm v,\ \ \nabla\cdot\bm v=0, \label{basic}
\end{eqnarray}
where $\bm v(\bm x, t)$ is the velocity field, $\nu$ is the (kinematic) viscosity and $P$ is the
fluid pressure divided by the density, $\nabla^2 P=-\nabla_i\nabla_j v_iv_j$. Among the central questions
posed by the NS equation is the existence of singularities in the solutions and the statistics
of the solutions in the limit of small $\nu$, both with and without forcing. Much light on these
questions could come from understanding the incompressible Euler equation obtained by
setting $\nu=0$ in Eq.~(\ref{basic}). However, the Euler equation is also ill-understood and questions
such as the existence of singularities are far from being settled.
In this letter we propose a new approach to the
equations, which we expect to be useful in tackling this old-standing problem.
We suggest that
relativistic conformal field theories (CFTs) provide a fundamental
microscopic viewpoint of the equations and the dynamics governed by them.

A complete (compressible) hydrodynamics is described by  five fields: the three velocity components, the temperature and the
particle density \cite{Frisch,Landau,Batchelor}. 
In a CFT there is no locally conserved charge corresponding to the particle density \cite{Landau,JeonYaffe}.
As a result, conformal hydrodynamics is described by
only four fields: the three velocity components and the temperature.
This implies that one cannot have a microscopic CFT structure for 
the complete hydrodynamics.
However,  the incompressible Euler and NS equations contain only three fields: the three velocity components, and may have
an underlying microscopic CFT structure. 

The hydrodynamics of relativistic conformal field theories is intrinsically relativistic as is
the microscopic dynamics. In particular, the non-relativistic limit
of a relativistic conformal hydrodynamics may not be well defined.
We will show that the limit of non-relativistic macroscopic motions of a CFT hydrodynamics is definable
and leads to the non-relativistic incompressible Euler
and NS equations for ideal and dissipative hydrodynamics of the CFT, respectively.
Thus, a relativistic conformal field theory contains the incompressible Euler and NS equations inside
it. We will explain the physical
reasons for the reduction and discuss the implications.

The hydrodynamics of relativistic conformal field theories has attracted much attention
recently in view of the AdS/CFT correspondence between gravitational theories on asymptotically
Anti-de-Sitter (AdS) spaces and CFTs \cite{Aharony:1999ti}.
Thus, for instance, a relation between black brane dynamics in  five-dimensional AdS space and fluid
dynamics on the four-dimensional boundary has been studied in \cite{Minwalla1,Minwalla2}.
Our results suggest that the incompressible Euler
and NS equations have a dual gravitational description.

Consider the hydrodynamical description of a relativistic CFT at finite temperature.
Hydrodynamics applies
under the condition that the correlation length of the fluid $l_{cor}$ is much smaller than the characteristic scale $L$ of variations of the macroscopic fields (such as moments of the stress-energy
tensor). In order to characterize the situation one introduces
the dimensionless Knudsen number $Kn\equiv l_{cor}/L$. Since the only dimensionfull parameter is the characteristic temperature of the fluid
$T$, one has by dimensional analysis,
\begin{eqnarray}&&
l_{cor}=(\hbar c/k_B T)G(\lambda),
\end{eqnarray}
where $\lambda$ denotes all the dimensionless parameters of the CFT. For example, one has $G\sim 1$ for
${\cal N}=4$ $SU(N)$ Super Yang-Mills (SYM) at large $N$ and large 't Hooft coupling, and more generally when a gravitational AdS dual description of a CFT is valid.

The stress-energy tensor of the CFT obeys
\begin{equation}
\partial_{\nu}T^{\mu\nu}=0,~~~~~T^{\mu}_{\mu}=0 .
\end{equation}
The
equations of relativistic hydrodynamics are determined by the constitutive relation expressing $T^{\mu\nu}$ in terms of the temperature $T(x)$ and the four-velocity field $u^{\mu}(x)$ satisfying $u_{\mu}u^{\mu}=-1$. Here $u^{\mu}$ and $T$ determine the local thermal equilibrium distribution of the fluid.
The constitutive relation has the form of a series in the small parameter $Kn\ll 1$,
\begin{eqnarray}&&
T^{\mu\nu}(x)=\sum_{l=0}^{\infty}T^{\mu\nu}_l(x),\ \ T^{\mu\nu}_l\sim (Kn)^l, \label{series}
\end{eqnarray}
where $T^{\mu\nu}_l(x)$ is determined by the local values of $u^{\mu}$ and $T$ and their
derivatives of a finite order. The smallness of $T^{\mu\nu}_l$ arises because it involves either the $l-$th derivative of $u^{\mu}$ or $T$ or because it contains the corresponding power of a lower-order derivative.
Keeping only the first term in the series gives ideal hydrodynamics within which the entropy
is conserved and the entropy density per unit volume $\sigma$ obeys conservation law $\partial_{\mu}(\sigma u^{\mu})=0$.
Dissipative hydrodynamics arises when
one keeps the first two terms in the series.

The ideal hydrodynamics
approximation for $T^{\mu\nu}$ does not contain
the spatial derivatives of the fields and would be the same for an infinite fluid with constant
$u^{\mu}$ and $T$. This allows to find the corresponding form of $T^{\mu\nu}$ easily \cite{Landau}.
First, we notice that in the rest frame of the fluid the stress-energy tensor
has the form (Pascal's law)
\begin{eqnarray}&&
T_{\mu\nu}=diag[e, p, p, p], \label{d2}
\end{eqnarray}
where $e$ is the energy density and $p$ is the pressure. The above equation and $T^{\mu}_{\mu}=0$
produce the equation of state $p=e/3$.
In particular, we see that in a CFT the speed of sound $c_s^2=c^2(\partial p/\partial e)_s$
is a constant, $c_s=c/\sqrt{3}$. The covariant expression for $T_{\mu\nu}$ reads
\begin{eqnarray}&&
T_{\mu\nu}=p\eta_{\mu\nu}+(e+p)u_{\mu}u_{\nu},
\label{d1}
\end{eqnarray}
and it gives the correct expression for $T_{\mu\nu}$ in the rest frame.
For a CFT we have, by dimensional analysis, $p=aT^4$ and $e=3aT^4$, where $T$
is the fluid temperature and $a$ is a constant, so that
\begin{eqnarray}&&
T_{\mu\nu}=aT^4\left[\eta_{\mu\nu}+4u_{\mu}u_{\nu}\right].
\end{eqnarray}
Note that $a$ enters as a multiplicative constant in the above expression and disappears
from the equations of motion $\partial_{\nu}T^{\mu\nu}=0$,
\begin{eqnarray}&&
\partial_{\nu}T^4\left[\eta^{\mu\nu}+4u^{\mu}u^{\nu}\right]=0. \label{eq1}
\end{eqnarray}
The resulting equation is completely independent of the microscopic structure of the theory - it
is the same for all conformal field theories.
It can be rewritten as \cite{Son}
\begin{eqnarray}&&
{\cal D} \xi=-\frac{1}{3}\partial_{\nu} u^{\nu},\ \ {\cal D}u^{\mu}=-\partial^{\mu}\xi+\frac{u^{\mu}\partial_{\nu} u^{\nu}}
{3}, \label{a1}
\end{eqnarray}
where $\xi\equiv \ln T$ and ${\cal D}=u^{\alpha}\partial_{\alpha}$.
The first equation follows from the second by multiplication
with $u_{\mu}$ and the use of $u^{\mu}u_{\mu}=-1$. This equation is equivalent
to the entropy conservation $\partial_{\mu}(\sigma u^{\mu})=0$ where
$\sigma=4aT^3$.

Consider now the non-relativistic slow motions limit $v\ll c$ where $\bm v$ is the three-velocity of the fluid. The latter is defined by $u^{\mu}=(\gamma, \gamma \bm v/c)$ and $\gamma=[1-v^2/c^2]^{-1/2}$.
Introducing the substantial derivative $D=\partial_t+\bm v\cdot \nabla$, see \cite{Landau}, we have
\begin{eqnarray}&&
{\cal D}=\frac{\gamma D}{c},\ \ \partial_{\nu}u^{\nu}=\frac{D \gamma}{c}+\frac{\gamma \nabla\cdot\bm v}{c},\ \
D\ln \gamma=\frac{\bm v D\bm v}{c^2-v^2}.\nonumber
\end{eqnarray}
Using the above identities one may rewrite Eqs.~(\ref{a1}) as
\begin{eqnarray}&&
\frac{\partial \xi}{\partial t}+\frac{2c^2}{3c^2-v^2}(\bm v\cdot\nabla)\xi=-\frac{c^2}{3c^2-v^2}\nabla\cdot\bm v,
\label{hyd1}\\&&
\frac{\partial v_i}{\partial t}+(\bm v\cdot\nabla) v_i=-(c^2-v^2)\left[\delta_{ij}-\frac{2v_iv_j}{3c^2-v^2}\right]
\nabla_j\xi\nonumber\\&&
+\frac{(c^2-v^2) v_i(\nabla\cdot\bm v)}{3c^2-v^2}. \label{hyd2}
\end{eqnarray}
Here and below latin indices will stand for spatial components.
The system in the above form allows to study the limit $v\ll c$ conveniently. In the lowest order in $v/c$
we find the equations of linearized hydrodynamics,
\begin{eqnarray}&&
\partial_t \xi=-\nabla\cdot\bm v/3,\ \ \partial_t \bm v=-c^2\nabla\xi. \label{lin}
\end{eqnarray}
The equations conserve the transversal component of velocity (just like ordinary linearized hydrodynamics equations do), while the rest of the variables
propagate as sound waves with the speed of sound $c_s=c/\sqrt{3}$, e. g.
\begin{eqnarray}&&
\partial_t^2 \xi=c_s^2\nabla^2 \xi.
\end{eqnarray}
Consider the leading order dynamics that follows from Eqs.~(\ref{hyd1})-(\ref{hyd2}) in the limit
$c\to\infty$. Note that linearized equations (\ref{lin}) have a well-defined limit $c\to\infty$ when
we introduce
\begin{eqnarray}&&
P=c^2\left[\xi+C\right],\ \ C=const,
\label{P}
\end{eqnarray}
and keep $P$ finite in the limit. This is the case in general. Eqs.~(\ref{hyd1})-(\ref{hyd2})
in terms of $P$ take the form
\begin{eqnarray}&&
\frac{1}{c^2}\left[
\frac{\partial P}{\partial t}+\frac{2c^2}{3c^2-v^2}(\bm v\cdot\nabla)P\right]=-\frac{c^2}{3c^2-v^2}\nabla\cdot\bm v,
\label{hyd3}\\&&
\frac{\partial v_i}{\partial t}+(\bm v\cdot\nabla) v_i=-\left(1-\frac{v^2}{c^2}\right)\left[\delta_{ij}-\frac{2v_iv_j}{3c^2-v^2}\right]
\nabla_jP\nonumber\\&&
+\frac{(c^2-v^2) v_i(\nabla\cdot\bm v)}{3c^2-v^2}. \label{hyd4}
\end{eqnarray}
In the limit $c\to\infty$ we obtain
\begin{eqnarray}&&
\frac{\partial \bm v}{\partial t}+(\bm v\cdot\nabla) \bm v=-\nabla P,\ \ \nabla\cdot\bm v=0, \label{Eul}
\end{eqnarray}
that is incompressible Euler equation.
The constant in the definition of $P$ remains undetermined as it does not influence the dynamics of $\bm v$ at all. The assumption that $P$ is finite at $c\to\infty$ is consistent with the relation $\nabla^2 P=-\nabla_j \nabla_i v_iv_j$ following from Eq.~(\ref{Eul}).

It is instructive to consider the above reduction in terms of the stress-energy tensor.
Introducing a new tensor ${\tilde T}^{\mu\nu}$ by ${\tilde T}^{00}=T^{00}$, ${\tilde T}^{0i}=cT^{0i}$ and ${\tilde T}^{ij}=c^2T^{ij}$, we may rewrite $\partial_{\nu}T^{\mu\nu}=0$ as
\begin{eqnarray}&&
\frac{\partial {\tilde T}^{\mu 0}}{\partial t}+\frac{\partial {\tilde T}^{\mu i} }{\partial x^i}=0. \label{tilde}
\end{eqnarray}
We omit the multiplicative constant $a$ and get
\begin{eqnarray}&&
{\tilde T}^{00}\approx 3T_0^4+O(v^2/c^2),\ \ {\tilde T}^{0i}=4T_0^4 v_i,\ \label{as1}\\&&
{\tilde T}^{ij}=c^2T_0^4\delta_{ij}
+4T_0^4\left[P\delta_{ij}+v_{i}v_{j}+O(v^2/c^2)\right], \label{as2}
\end{eqnarray}
where
\begin{eqnarray}&&
T=T_0\left[1+P/c^2+o(1/c)\right],\ \ T_0\equiv\exp[-C].
\label{expan}
\end{eqnarray}
It is easy to see that Eqs.~(\ref{tilde})-(\ref{as2}) reproduce the incompressible Euler equation. Note that the temperature and thus the pressure are
almost homogeneous in the limit. The physical reason for this is that here the limit $v/c\to 0$ also
corresponds to the limit of small Mach number. The Mach number is defined as the ratio of the characteristic velocity
of the fluid to the speed of sound $c_s$ \cite{Landau}. In this limit the sound propagation is instantaneous, which leads to instantaneous homogenization of the pressure.  In the case of a  CFT,
where there is only one independent thermodynamic variable, such homogenization implies incompressibility
as in many other cases of small Mach number \cite{Landau}.
Thus, the ideal hydrodynamics of a general CFT gives the incompressible Euler equation in the limit of
non-relativistic velocities.

Consider next the dissipative hydrodynamics obtained by keeping $l=1$ term in the
series in Eq.~(\ref{series}). In the Landau frame
\cite{Landau,Son} the stress-energy tensor reads
\begin{eqnarray}&&
T_{\mu\nu}=aT^4\left[\eta_{\mu\nu}+4u_{\mu}u_{\nu}\right]-c\eta\sigma_{\mu\nu},
\end{eqnarray}
where $\sigma_{\mu\nu}$ obeys $\sigma_{\mu\nu}u^{\nu}=0$ and is given by
\begin{eqnarray}&&
\sigma_{\mu\nu}=\left(\partial_{\mu} u_{\nu}+
\partial_{\nu} u_{\mu}+u_{\nu}u^{\rho}\partial_{\rho} u_{\mu}+
u_{\mu}u^{\rho}\partial_{\rho} u_{\nu}\right) \nonumber\\&&
-\frac{2}{3}\partial_{\alpha}u^{\alpha}
\left[\eta_{\mu\nu}+u_{\mu}u_{\nu}\right]. \label{str}
\end{eqnarray}
Note that dissipative hydrodynamics of a CFT is determined by only one kinetic coefficient -
the shear viscosity $\eta$.
The bulk viscosity $\zeta$ vanishes for the CFT, while the absence
of the particle number conservation and the use of the Landau frame allow to avoid the use of
heat conductivity (which is not an independent coefficient here \cite{JeonYaffe}).
The dissipative hydrodynamics of different CFTs differs by the value of the dimensionless function
$F(\lambda)$ in
\begin{eqnarray}&&
\eta=aF(\lambda)T^3,
\end{eqnarray}
(here and below we put $\hbar=k_B=1$ until stated otherwise).
For example, for strongly coupled CFTs described by an AdS gravity dual,
$\eta$
can be determined from the universal ratio $\eta/\sigma=1/4\pi$ \cite{PSS},
giving $F=1/\pi$.

Dropping a multiplicative constant in $T_{\mu\nu}$
we have
\begin{eqnarray}&&
T_{\mu\nu}=T^4\left[\eta_{\mu\nu}+4u_{\mu}u_{\nu}\right]-c F(\lambda) T^3 \sigma_{\mu\nu}.
\end{eqnarray}
The tensor $\sigma_{\mu\nu}$ contains both space and time derivatives, and as a result $\partial_{\nu}T^{\mu\nu}=0$ contains second derivatives both in space and time.
Thus, it does not lead to an equation that is first order in time.
Note, however, that the time-derivatives in $\sigma_{\mu\nu}$ can be substituted by their
ideal hydrodynamics expressions. Indeed, within hydrodynamics the time derivatives of fields are given
by a series in the spatial gradients of the fields, while second derivative terms in $\sigma_{\mu\nu}$
would correspond to a higher order of expansion than the considered one, cf. \cite{Weinberg}. The use of Eqs.~(\ref{a1})
allows to rewrite Eq.~(\ref{str}) as
\begin{eqnarray}&&
\sigma_{\mu\nu}=\partial_{\mu}u_{\nu}+\partial_{\nu}u_{\mu}-u_{\nu}\partial_{\mu}\xi-u_{\mu}\partial_{\nu}\xi -2\eta_{\mu\nu}\partial_{\alpha}u^{\alpha}/3.\nonumber
\end{eqnarray}
The generalization of Eqs.~(\ref{a1}) which includes the viscous contribution to the stress-energy tensor is
\begin{eqnarray}&&
{\cal D} \xi=-\frac{\partial_{\nu} u^{\nu}}{3}+\frac{c F(\lambda) I}{12T}, \ \ \
I\equiv \sigma^{\mu\nu}\partial_{\nu}u_{\mu}, \nonumber\\&& {\cal D}u^{\mu}=-\partial^{\mu}\xi+\frac{u^{\mu}\partial_{\nu} u^{\nu}}{3}
-\frac{cFIu^{\mu}}{3T}+\frac{cFA^{\mu}}{4T}, \label{v2}\\&&
A^{\mu}\equiv3\sigma^{\mu\nu}\partial_{\nu}\xi+\partial_{\nu}\sigma^{\mu\nu}=
\partial_{\nu}\left[\sigma^{\mu\nu}T^3 \right]/T^3.
\end{eqnarray}
Using Eqs.~(\ref{a1}) one can write
\begin{eqnarray}&&
I
=(\partial_{\nu}u_{\mu})\left[\partial^{\mu}u^{\nu}+\partial^{\nu}u^{\mu}\right]
+(\partial^{\mu}\xi)(\partial_{\mu}\xi)
-5(\partial_{\nu}u^{\nu})^2/9. \nonumber
\end{eqnarray}
Passing from $u^{\mu}$ to $\bm v$ as
in Eqs.~(\ref{hyd1})-(\ref{hyd2}) one finds
\begin{eqnarray}&&
\frac{\partial \xi}{\partial t}+\frac{2c^2}{3c^2-v^2}(\bm v\cdot\nabla)\xi=-\frac{c^2}{3c^2-v^2}\nabla\cdot\bm v
\nonumber\\&& +\frac{c^2F(\lambda)(c^2+v^2)I}{4T\gamma(3c^2-v^2)}
-\frac{cF(\lambda)(c^2-v^2)(\bm v\cdot\bm A)}{4T(3c^2-v^2)}, \label{hyd3}
\\&&
\frac{\partial v_i}{\partial t}\!+\!(\bm v\cdot\nabla) v_i=\frac{(c^2\!-\!v^2) v_i(\nabla\cdot\bm v)}{3c^2\!-\!v^2}
\!-\!\frac{(c^2\!-\!v^2) v_ic^2FI}{(3c^2\!-\!v^2)T\gamma}\nonumber\\&&
-(c^2-v^2)\left[\delta_{ij}-\frac{2v_iv_j}{3c^2-v^2}\right]
\left(\nabla_j\xi-\frac{cF(\lambda)A_j}{4T}\right)
. \label{hyd4}
\end{eqnarray}
To rewrite $A_i$ in terms of $\bm v$
and $\xi$ one can write $A_i$ as
\begin{eqnarray}&&
A_i=2\partial_i (\partial_{\nu}u^{\nu})/3+4(\partial_iu^{\nu})(\partial_{\nu}\xi)+\partial_{\nu}\partial^{\nu}u_i
-u_i\partial_{\nu}\partial^{\nu}\xi
\nonumber\\&&+2
(\partial_{\nu}u_i)(\partial^{\nu}\xi)-3u_i(\partial^{\nu}\xi)
(\partial_{\nu}\xi)-2(\partial_{\nu}u^{\nu})(\partial_i\xi). \label{sum}
\end{eqnarray}
Further, the terms in the above sum can be fixed using Eqs.~(\ref{a1}).
We are interested in the leading order
dynamics in the limit $c\to \infty$. In this limit, remembering that the time-derivatives of $\bm v$ are finite,
and using the definition of $P$ (\ref{P}), where derivatives of $P$ are finite (see below), one finds that $A_i$ is
of order $1/c$. The only terms in the sum in Eq.~(\ref{sum}) which are of this order are
\begin{eqnarray}&&
\partial_i (\partial_{\nu}u^{\nu})\approx \partial_i\nabla\cdot\bm v/c,\ \ \partial_{\nu}\partial^{\nu}u_i\approx \nabla^2v_i/c,
\end{eqnarray}
while the rest of the terms are of higher order.
We find
\begin{eqnarray}&&
A_i\approx 2\partial_i\nabla\cdot\bm v/3c+\nabla^2v_i/c.\label{Avec}
\end{eqnarray}
Analogously, after a straightforward but lengthy calculation one finds
\begin{eqnarray}&&
I\approx \left[\partial_j v_i+\partial_iv_j\right]\partial_iv_j/c^2
-2[\nabla\cdot\bm v]^2/3c^2.
\end{eqnarray}
Thus, in the leading order in $1/c$, Eq.~(\ref{hyd3}) becomes
\begin{eqnarray}&&
\nabla\cdot\bm v=(F(\lambda)/4T)\Bigl[\left(\partial_j v_i+\partial_iv_j\right)\partial_iv_j-2[\nabla\cdot\bm v]^2/3
\nonumber\\&&-\bm v\nabla^2\bm v
-2(\bm v\cdot\nabla)\nabla\cdot\bm v/3\Bigr]. \label{appr}
\end{eqnarray}
Restoring the dimensions, one finds that the ratio of the RHS to the LHS is governed by dimensionless parameter
\begin{eqnarray}&&
S\equiv \frac{\hbar F(\lambda) V}{Lk_BT}\sim Kn\frac{V}{c}\frac{F(\lambda)}{G(\lambda)}.
\end{eqnarray}
Here $V$ is the characteristic value of the velocity, $L$ is the characteristic scale and $T$ is the characteristic temperature. 
Generally $\eta$ obeys the estimate $\eta\sim (e/c^2)l_{cor}c_s$, see \cite{FBB}, implying $F(\lambda)\sim G(\lambda)$.
Then $S\sim Kn (V/c)\ll 1$.
In special
situations where $F/G$ is not of order one (cf. \cite{FBB}), at sufficiently small $V/C$ and $Kn$ we
still have $S\ll 1$ and Eq.~(\ref{appr}) reduces to just
\begin{eqnarray}&&
\nabla\cdot\bm v=0.
\end{eqnarray}
Using the above one finds for Eq.~(\ref{hyd4})
\begin{eqnarray}&&
\frac{\partial \bm v}{\partial t}+(\bm v\cdot\nabla)\bm v=-\nabla P+\nu\nabla^2\bm v,\ \
\nu\equiv\frac{\hbar c^2 F(\lambda)}{4 k_B T_0},\nonumber
\end{eqnarray}
where the constant $T_0$ is still determined by Eq.~(\ref{expan}). The last term in the first line of Eq.~(\ref{hyd4}) is smaller than $\nu\nabla^2\bm v$ by a factor $V^2/c^2\ll 1$ and can be omitted.

Thus, in the limit $v/c\to 0$ dissipative hydrodynamics of the CFT produces
the incompressible NS equation with the effective viscosity determined by the almost homogeneous
background of the temperature (corrections to which are determined by non-relativistic "pressure" $P$).
The underlying physics is the same as for the ideal hydrodynamics.
In terms of the stress-energy tensor, Eq.~(\ref{as1})
is unchanged
while Eqs.~(\ref{as2}) is modified
to
\begin{eqnarray}&&
{\tilde T}^{ij}=c^2T_0^4\delta_{ij}
+4T_0^4\left[P\delta_{ij}+v_{i}v_{j}+O(v^2/c^2)\right]\nonumber\\&&
-\frac{\eta c^2}{a}\left[\partial_jv_i+\partial_iv_j-\frac{2}{3}\delta_{ij}\nabla\cdot\bm v\right]. \label{as4}
\end{eqnarray}
The corrections however are now governed not only by $v/c$ but also by $S$ - the expansion is now
two-parametric.
The viscosity $\nu$ obeys $\nu\sim (F/G)l_{cor}c_s$. For $F\sim 1$, at room temperature it
is of order $10^8 cm^2/s$ which is about ten orders of magnitude larger than the one of air. In ordinary circumstances such fluid would behave almost
as a solid. On the other hand, $\nu$ falls linearly with temperature and for very high temperatures, e.g. of strongly coupled thermal plasma,
a fluid-like behavior will be displayed. The relevant dimensionless parameter that characterizes
the viscosity is the Reynolds number $Re\equiv V L/\nu$ and it can take an arbitrary value in accord with
the external conditions.

We found that the incompressible Euler and Navier Stokes equations are contained in relativistic
conformal field theories at finite temperature, and are obtained in the limit of slow motions.
In view of that, we are lead
to propose that relativistic
conformal field theories can provide a fundamental
microscopic viewpoint of the Euler and NS equations and the dynamics governed by them.
Thus, we expect that one can study the Euler and NS equations, their symmetries and their dynamics
using tools of relativistic CFTs. For instance, the anomalous scaling exponents of turbulence \cite{Frisch,review} should be encoded in the relativistic CFTs dynamics.
The existence of an AdS gravity description  for a large class of the CFTs, implies the existence
of a gravity picture for the Euler and NS equations.
Having a gravity description can provide an insight to basic issues in turbulence
such as the existence of singularities in the solutions.
Recently, some features of $2+1$ dimensional incompressible turbulence were shown numerically
to display conformal invariance \cite{Grisha1,Grisha2}. Theoretical reasons for the finding are not
clear yet. We hope that our result (which generalizes to an arbitrary space dimension straightforwardly)
may be of help to shed light on this discovery.

We would like thank G. Falkovich, S. Theisen and S. Yankielowicz for valuable discussions.
This work is supported in part by the Israeli Science Foundation center of excellence,
by the Deutsch-Israelische Projektkooperation, by the the US-Israel binational science foundation and
by the European Network.


\begin{thebibliography} {99}

\bibitem{Frisch} U. Frisch, \textit{Turbulence: The Legacy of A. N.
Kolmogorov}, Cambridge University Press, 1995.

\bibitem{Landau} L. D. Landau and E. M. Lifshitz, \textit{Fluid Mechanics}  (Butterworth-Heinemann, 2000).

\bibitem{Batchelor} G. K. Batchelor, \textit{An introduction to fluid dynamics},
Cambridge University Press, 1967.

\bibitem{JeonYaffe} S. Jeon and L. G. Yaffe, Phys. Rev. D \textbf{53} (1996) 5799.

\bibitem{Aharony:1999ti}
  O.~Aharony, S.~S.~Gubser, J.~M.~Maldacena, H.~Ooguri and Y.~Oz,
  Phys. Rept.  {\bf 323} (2000) 183.


\bibitem{Minwalla1} S. Bhattacharyya, S. Minwalla, V. E. Hubeny and M. Rangamani, JHEP 02 (2008) 045.

\bibitem{Minwalla2} S. Bhattacharyya, V. E. Hubeny, R. Loganayagam, G. Mandal, S. Minwalla, T. Morita,
M. Rangamani and H. Reall, JHEP 06 (2008) 055.




\bibitem{Son}  R. Baier, P. Romatschke, D. T. Son, A. O. Starinets and M. A. Stephanov,  arXiv:0712.2451.






\bibitem{PSS} G. Policastro, D. T. Son, and A. O. Starinets, Phys. Rev. Lett. \textbf{87} (2001) 081601.

\bibitem{Weinberg} S. Weinberg, Astrophys. J. \textbf{168} (1971) 175.

\bibitem{FBB} I. Fouxon, G. Betschart, and J. D. Bekenstein, Phys. Rev. D \textbf{77} (2008) 024016.

\bibitem{review} G. Falkovich, K. Gawedzki, and M. Vergassola, Rev. Mod. Phys. \textbf{73} (2001) 913.

\bibitem{Grisha1} D. Bernard, G. Boffetta, A. Celani, and G. Falkovich, Nature Physics \textbf{2} (2006)
124.

\bibitem{Grisha2} D. Bernard, G. Boffetta, A. Celani, and G. Falkovich, Phys. Rev. Lett. \textbf{98} (2007)
024501.





\end{thebibliography}
\end{document}